\documentclass[twocolumn,english,aps,showpacs,nofootinbib]{revtex4}
\usepackage[T1]{fontenc}
\usepackage[latin9]{inputenc}
\usepackage{amsmath}
\usepackage{amssymb}
\usepackage{esint}
\usepackage{graphicx}
\makeatletter
\@ifundefined{textcolor}{}
{%
 \definecolor{BLACK}{gray}{0}
 \definecolor{WHITE}{gray}{1}
 \definecolor{RED}{rgb}{1,0,0}
 \definecolor{GREEN}{rgb}{0,1,0}
 \definecolor{BLUE}{rgb}{0,0,1}
 \definecolor{CYAN}{cmyk}{1,0,0,0}
 \definecolor{MAGENTA}{cmyk}{0,1,0,0}
 \definecolor{YELLOW}{cmyk}{0,0,1,0}
 }

\usepackage{nicefrac}\usepackage{dsfont}
\usepackage[normalem]{ulem}\@ifundefined{definecolor}
 {\usepackage{color}}{}

\usepackage{enumerate}

\makeatother

\usepackage{babel}
\begin{document}

\title{Driven diffusive systems with mutually interactive Langmuir kinetics}
\author{H. D. Vuijk}
\author{R. Rens}
\author{M. Vahabi}
\author{F. C. MacKintosh}
\author{A. Sharma}
\affiliation{Department of Physics and Astronomy, VU University, Amsterdam, The Netherlands}


\date{\today}
\begin{abstract}
We investigate the simple one-dimensional driven model, the totally asymmetric exclusion process, coupled to mutually interactive Langmuir kinetics. This model is motivated by recent studies on clustering of motor proteins on microtubules. In the proposed model, the attachment and detachment rates of a particle are modified depending upon the occupancy of neighbouring sites. We first obtain continuum mean-field equations and in certain limiting cases obtain analytic solutions. We show how mutual interactions increase (decrease) the effects of boundaries on the phase behavior of the model. We perform Monte Carlo simulations and demonstrate that our analytical approximations are in good agreement with the numerics over a wide range of model parameters. We present phase diagrams over a selective range of parameters.
\end{abstract}

\pacs{87.16.Uv, 05.70.Ln, 87.10.Hk}
\maketitle


\section{introduction}
Driven diffusive systems show a very rich behavior. Even one-dimensional systems exhibit boundary induced phase transitions\cite{krug1991boundary, evans1995spontaneous, evans1998phase, kafri2002criterion, parmeggiani2003phase, parmeggiani2004totally} with a complex phase behavior. One such model is that of a totally asymmetric exclusion process (TASEP)\cite{spohn1991large,privman2005nonequilibrium,cates2010soft,domb2000phase} coupled to the Langmuir kinetics (LK)\cite{parmeggiani2004totally}. In that model a single species of particles performs unidirectional hopping on a 1D lattice. The particles are assumed to have hard-core repulsion which prevents more than one particle from occupying the same lattice site. Such a system is coupled to Langmuir kinetics by allowing for adsorption (desorption) of particles at an empty (filled) lattice site with fixed respective kinetic rates. It was shown in Refs.~\cite{parmeggiani2003phase,parmeggiani2004totally} that the combination of TASEP and LK results in nonconserved dynamics with unusual features such as the appearnce of a high-low coexistence phase separated by stable discontinuities in the density profile. The novel phase behavior has its origin in the competing kinetics of TASEP and LK. However, in the thermodynamic limit, it is expected that the bulk effects are predominant with boundaries becoming insignificant. In fact, the competition between bulk and boundary dynamics can occur only if one rescales the attachment (detachment) kinetic rates \cite{parmeggiani2003phase,parmeggiani2004totally} such that they decrease with increasing system size in a particular fashion.

Besides being fundamentally interesting, understanding nonequilibrium physics of driven systems is of particular interest in biological systems\cite{macdonald1968kinetics,parmeggiani2004totally}. One such particular system is that of molecular motors performing directed motion along one-dimensional molecular tracks. Typically kinetic rates are such that the fraction of track over which the motor moves before detaching is finite\cite{howard2001mechanics}. This allows for the bulk dynamics to compete with the boundary, potentially giving rise to unusual nonequilibrium stationary states. Recently, exclusion process on networks have been used to model cytoskeletal transport\cite{neri2013exclusion}. It was shown that active transport processes spontaneously develop density heterogeneities at various scales. An important aspect that needs to be included in a study of motor transport is that of mutual interactions (MI) between motors. Seitz \emph{et al}\cite{seitz2006processive} observed that in presence of an obstacle, a molecular motor walking on a microtubule tends to stay attached for a longer time. Muto \emph{et al}\cite{muto2005long} reported on long-range cooperative binding of kinesin to a microtubule. The detachment could depend on the biochemical state of the motor\cite{crevel2004kinesin,telley2009obstacles} which might itself be determined by the presence or absence of neighbouring motors.

In a recent study on kinesin-1 motors moving on microtubule\cite{roos2008dynamic}, the authors performed numerical simulations of binding/unbinding dynamics incorporating mutually attractive interaction between the motor proteins. Their results were in agreement with the experimental observation, in particular clustering of motors on microtubules. Mututal interactions in addition to the hard-core repulsion introduce additional correlations as in the Katz-Lebowitz-Spohn (KLS) model, which is a generic model of interacting driven diffusive systems\cite{katz34nonequilibrium}. By modifying the hopping rate of particles depending upon the occupancy of next nearest neighbour, the model gives rise to exotic features such as localized downwards shocks and phase separation into three distinct regimes\cite{popkov2003localization}. However,  in the case of molecular motors, due to the mutual interactions, the attachment and detachment rates of a motor molecule are modified depending upon the state of the neighbouring sites\cite{roos2008dynamic}. Assuming that the hopping rate is unaltered, this corresponds to the ordinary TASEP (with no correlations besides the hard-core repulsion) with density (local) dependent LK.

In this paper, we focus on the TASEP coupled to mutually interactive Langmuir kinetics. We investigate how mutual interactions can tilt the balance in favor of predominantly bulk effects by enhancing LK. We show that that this is indeed the case when both the attachment and detachment rates are enhanced significantly due to the mutual interactions. In the case of the kinetic rates being significantly reduced due to the interactions, one suppresses the bulk effects giving rise to rich and complex phase behavior. We also explore the more interesting scenario in which the kinetic rates are modified in an asymmetric fashion. The paper is organized in the following way. In Sec.~\ref{Model}, we present the model composed of TASEP coupled to the modified Langmuir Kinetics. We first present the modfication to the LK due to the mutual interactions. We then obtain continuum mean field equation describing the steady state density profile of particles on the lattice. In Sec.~\ref{modifiedLK} we study three different cases of modified LK. We first consider the case in which the unmodified LK rates are assumed to be equal and mutual interaction enhances (suppresses) the attachment and detachment rates by the same factor. The second case corresponds to the unmodified LK rates being equal, but the mutual interaction enhances one and suppresses the other by the same factor. The last case is the most general one in which all the model parameters are freely chosen. We do not explore this case in detail. In Sec.~\ref{conclusion}, we summarize our findings.


\section{The Model} \label{Model}

\begin{figure}
\includegraphics[width = \columnwidth]{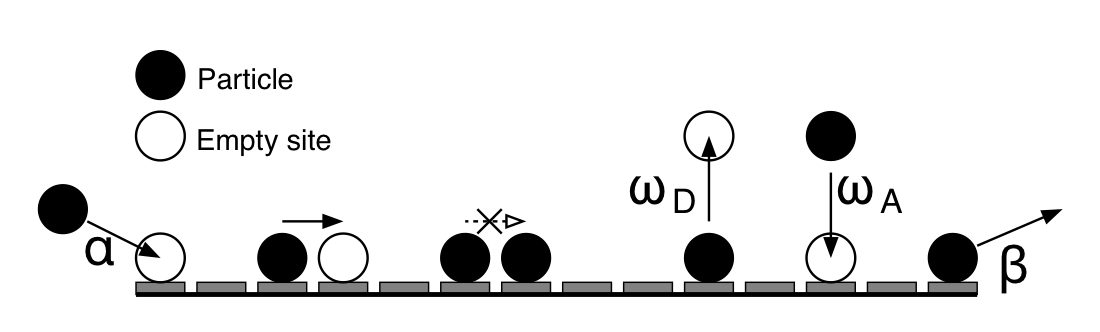}
\caption{
A graphic representation of the TASEP model with Langmuir dynamics. Particles are injected at the first site with rate $\alpha$ and extracted at the last site with rate $\beta$.The particles move exclusively to the right, with unitary rate, if the next site is empty. In this model the only interaction between the particles is the hard-core repulsion, this means that hopping over particles and multiple occupation are not allowed. The injection, extraction and hopping to the next site constitute the TASEP model. The Langmuir dynamics consists of the detachment and attachment from and to the background. The attachment and detachment rates are $\omega_{A}$ and $\omega_{D}$ respectively. For analysis of this model see for example Refs.~\citep{parmeggiani2004totally,parmeggiani2003phase}. Mutual interactions are incorporated by modifying the attachment and detachment rates; see Fig.~\ref{mirates}.
}
\label{tasep_lk_model}
\end{figure}

The model consists of a 1D lattice with sites $i=1,2, ... ,N$; see Fig.  \ref{tasep_lk_model}.
Each site on the lattice can either be occupied by one particle or no particle. There are three different sub-processes that govern the dynamics of the system:
\begin{enumerate}[(a)]
\item Particles are injected at site $1$ with rate $\alpha$ and extracted at site $N$ with rate $\beta$.
\item Particles at site $i=1,...N-1$ can hop to site $i+1$ if site $i+1$ is unoccupied.
\item Particles at site $i=2,...,N-1$ can detach from the lattice with rate $\omega_D$ and attach to site $i=2,...,N-1$ with rate $\omega_A$.
\end{enumerate}
All the rates are defined such that the hopping rate is unitary. Processes $a$ and $b$ are the dynamics of the TASEP model, process $c$ is the interaction with the background. The interaction with the background is called Langmuir kinetics. We note that the only interaction between the particles is assumed to be the hard-core repulsion.

For the sake of completion we show the equations for site-occupancy below. These equations are the same as reported in Ref.~\cite{parmeggiani2003phase}. The equation for the occupancy of each site is given by
\begin{multline}\label{rate eq tasep lk ma bulk}
\frac{\textit{d}}{\textit{d}t} n_i(t) = n_{i-1}(t)[1-n_i(t)]-n_i(t)[1-n_{i+1}(t)] \\ + \omega_A[1-n_i(t)] - \omega_D n_i(t),
\end{multline}
with $n_i$ the occupancy at site $i$, which can either be one or zero. The equations for the boundary sites are
\begin{equation}\label{rate eq tasep lk ma i=1}
\frac{\textit{d}}{\textit{d}t} n_1(t) = \alpha [1-n_1(t)] - n_1(t)[1-n_2(t)],
\end{equation}
\begin{equation}\label{rate eq tasep lk ma i=N}
\frac{\textit{d}}{\textit{d}t} n_N(t) = n_{N-1}(t)[1-n_N(t)]-\beta n_N(t).
\end{equation}

\begin{figure}
\includegraphics[width = \columnwidth]{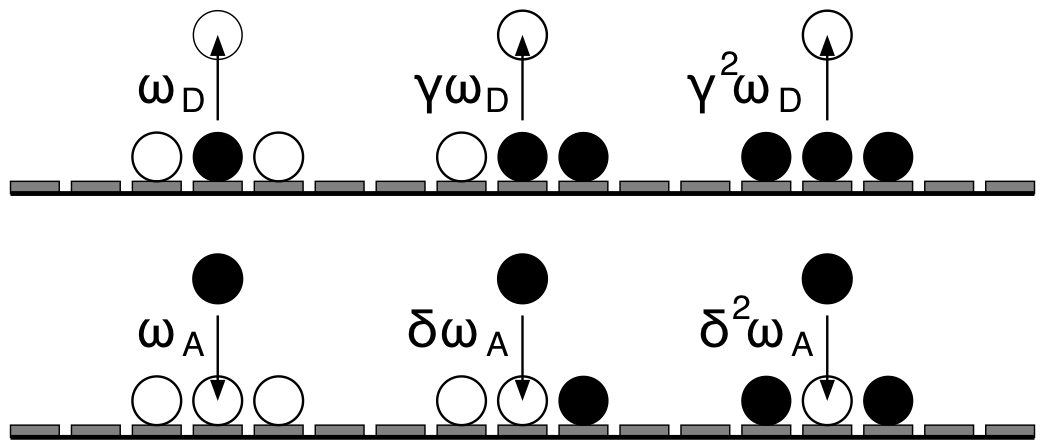}
\caption{
The mutual interaction are incorporated in the TASEP model with LK dynamics by modifying the attachment and detachment rates if neighbouring sites are occupied. For each occupied neighbouring site the attachment rate is multiplied by $\delta$ and the detachment rate is multiplied by $\gamma$.
}
\label{mirates}
\end{figure}

The mutual interactions of the particles are included in the equation by modifying the attachment and detachment rates $\omega_A$ and $\omega_D$ respectively. The attachment (detachment) rate if both neighbouring sites are unoccupied is $\omega_A$ ($\omega_D$). If either the left or right neighbouring site is occupied the attachment (detachment) rate becomes $\delta\omega_A$ ($\gamma\omega_D$), and if both neighbouring sites are occupied $\delta^2\omega_A$ ($\gamma^2\omega_D$); see Fig.~\ref{mirates}. In our model, the hopping rate of particles on the lattice is unaltered in presence of mutual interactions. This is in contrast with the KLS model where the hopping rates are modified according to the occupancy of nearest and next-nearest neighbours\cite{katz34nonequilibrium} whereas the binding/unbinding kinetics remain unaltered.

In order to include the mutual interactions of the particles the following substitutions are needed:
\begin{equation}\label{omega_A}
\omega_A \rightarrow \omega_A[1+(n_{i+1}+n_{i-1})(\delta-1)+n_{i+1}n_{i-1}(\delta^2-2\delta+1)],
\end{equation}
\begin{equation}\label{omega_D}
\omega_D \rightarrow \omega_D[1+(n_{i+1}+n_{i-1})(\gamma-1)+n_{i+1}n_{i-1}(\gamma^2-2\gamma+1)].
\end{equation}

At present it is not clear how already bound motors modify the binding kinetics of motors. It has been suggested that presence of a bound motor could change the lattice locally somehow leading to a modified binding/unbinding kinetics\cite{krebs2004complex}. However, the underlying mechanism remains unkown.

In order to obtain useful solutions for the distribution of particles on the lattice, the following two steps are required. First one goes from the equation for the occupation of the sites, where each site can have either value one or zero, to an equation of the average occupation of the sites. Second in the limit of large system sizes a semi-continuous variable $x$ instead of the discrete parameter $i$ is used for the position on the lattice. This method is the same as used in \citep{parmeggiani2004totally}.
The average density at a site is defined as $\langle n_i(t)\rangle \equiv \rho_i(t)$. In stationary state the average $\langle n_i \rangle$ can either be a time or a sample average. In order to take the averages of Eqs.~(\ref{rate eq tasep lk ma bulk}, \ref{rate eq tasep lk ma i=1}, \ref{rate eq tasep lk ma i=N}) with substitutions (\ref{omega_A}), (\ref{omega_D} the higher order correlations are needed. Instead of solving these equations exactly a mean field approach is used which consists of the approximation:
\begin{equation}\label{random_phase_approx}
\langle n_i(t)n_{i+1}(t) \rangle \approx \langle n_i(t)\rangle \langle n_{i+1}(t) \rangle.
\end{equation}

The lattice constant $\epsilon$ is defined as $\epsilon \equiv L/N$. For simplicity the length of the lattice is fixed to one $L=1$. For large system sizes, $N\gg 1$ the quasicontinuous position variable $x=i/N$ is introduced. This means that the average density at site $i$ is now defined as $\langle n_i(t) \rangle \equiv \rho(x,t)$.
The equation for the average density profile in stationary state, to leading order in $\epsilon$ becomes
\begin{multline}\label{mi_equation_semi_continuous}
0=\frac{\epsilon}{2} \partial_x^2 \rho + (2\rho-1)\partial_x \rho + \Omega_A[1+\rho(\delta-1)]^2 (1-\rho) \\- \Omega_D[1+\rho(\gamma-1)]^2 \rho.
\end{multline}
The $[1+\rho(\delta-1)]^2$ and $[1+\rho(\gamma-1)]^2$ parts of the equation are due to the mutual interactions. In equation \ref{mi_equation_semi_continuous} the total detachment/attachment rates are used, defined as $\Omega_A=\omega_A N$ and $\Omega_D=\omega_D N$.
The equations for the boundary sites (Eqs.~(\ref{rate eq tasep lk ma i=1}, \ref{rate eq tasep lk ma i=N})) become the boundary conditions.
 \begin{equation}\label{boundary conditions}
 \rho(0)=\alpha\ , \ \ \ \rho(1)=1-\beta.
 \end{equation}

 One can now take the continuous limit, $\epsilon \rightarrow 0$ for a normalized lattice this means $N\rightarrow\infty$. In order to ensure that the attachment/detachment rates per unit length do not become infinite, the total rates $\Omega_A$ and $\Omega_D$ are kept constant. In the continuous limit, $\epsilon\rightarrow 0$, the second order differential equation (\ref{mi_equation_semi_continuous}) becomes a first order differential equation, but the two boundary conditions remain. This means that the problem is overdetermined. However one can find solutions to the equation in the continuum limit that satisfy one of the two boundary conditions. The full density profile is is constructed from the possible solutions. The crossover position from one solution to an other is obtained by matching the currents $j(x)$ \citep{parmeggiani2004totally}.
\begin{equation}\label{curent}
j(x)=\rho(x)[1-\rho(x)]
\end{equation}
 For a normalized lattice in the continuous limit the crossover region is localized and a discontinuity in the density profile appears. Though the crossover region in this case is localized it does span a finite number of sites implying that in the case of a finite sized lattice the crossover region spans a finite fraction of the normalized lattice.

The model without MI exhibits a particle-hole symmetry, in the sense that a particle attaching to the lattice means that a vacancy detaches from the lattice and vice versa. The same holds for a particle entering (leaving) the system on the first (last) site, which can be seen as a vacancy leaving (entering) the system. And a particle hooping to a neighbouring site on the right equals a vacancy hopping to the left. Due to this symmetry the following transformations
\begin{subequations}
\begin{align}
n_i(t) & \leftrightarrow 1-n_{N-i}(t),\\
\alpha & \leftrightarrow \beta,\\
\omega_A & \leftrightarrow \omega_D,
\end{align}
\end{subequations}
leave Eqs.~(\ref{rate eq tasep lk ma bulk}), (\ref{rate eq tasep lk ma i=1}) and (\ref{rate eq tasep lk ma i=N}) invariant \citep{parmeggiani2004totally}. However this particle-hole symmetry is no longer apparent if MI is included.

\section{Modified Langmuir Kinetics}\label{modifiedLK}

In this section the solutions of Eqn.~(\ref{mi_equation_semi_continuous}) in the continuum limit are presented for three different cases. The first case corresponds to where LK rates are both enhanced or reduced simultaneously by the same amount, and second case to where $\Omega_{A}$ is enhanced while $\Omega_{D}$ is reduced by the same amount. The third case is the most general one in which the attachment and detachment rates are independently modified. This case ix explored in least details. With these solutions the density profiles are constructed and compared with Monte Carlo simulations of the model. Besides the density profiles phase diagrams are made which show the characteristics of the solutions for different values of $\alpha$ and $\beta$.


\subsection*{Case 1: mutual interaction with enhanced LK rates}

The model simplifies significantly in the case that $\Omega_A=\Omega_D \equiv \Omega$ and $\delta=\gamma \equiv 1+\eta$. This means that the attachment and detachment rates are both multiplied by $1+\eta$ for each occupied neighbouring site. Values of $\eta<-1$ result in negative rates, therefore $\eta$ is restricted to values larger than -1. Positive $\eta$ increases and negative $\eta$ decreases the LK dynamics if neighbouring sites are occupied.
In this case Eqn.~(\ref{mi_equation_semi_continuous}) in the continuous limit becomes
\begin{equation}\label{density case1}
0=(2\rho -1) \left( \partial_x \rho - \Omega [1+\rho \eta]^2 \right).
\end{equation}
The special case where $\eta=0$, the case without mutual interactions, corresponds to the symmetric case analysed in \citep{parmeggiani2004totally}.
Eqn.~(\ref{density case1}) has three solutions. A constant solution $\rho_l=1/2$ which is the Langmuir isotherm. The other solutions are $\rho_{\alpha}$ and $\rho_{\beta}$ which obey the left and right boundary condition respectively.

\begin{equation}\label{rho alpha}
\rho_\alpha= \frac{\alpha + (1+ \eta \alpha) \Omega x}{1-(1+\eta \alpha)\eta \Omega x}
\end{equation}
\begin{equation}\label{rho beta}
\rho_\beta=\frac{1 -\beta + (\eta(1 - \beta) + 1) \Omega(x-1)}{1-(\eta(1 - \beta) +1) \eta \Omega (x-1)}
\end{equation}
The full density profile $\rho(x)$ is a combination of these solutions
\begin{equation}\label{rho xa xb case1}
\rho(x) = \begin{cases}
\rho_{\alpha}  & \text{for $0 \leq x \leq x_{\alpha}$},\\
\rho_l & \text{for $x_{\alpha} \leq  x \leq x_{\beta}$},\\
\rho_\beta  & \text{for $ x_{\beta} \leq x \leq 1$}.
\end{cases}
\end{equation}
Where $x_{\alpha}$ and $x_{\beta}$ are obtained by equating the currents of the solutions,  $j_\alpha(x_\alpha)=j_l$ and $j_\beta(x_\beta)=j_l$ \citep{parmeggiani2004totally}. The domain of the solutions can be explained as an attraction of the density to the Langmuir isotherm as one moves away from the boundary. As one moves away from the boundary the influence of the bulk dynamics i.e. the Langmuir kinetics becomes more dominant and therefore the density tries to reach the Langmuir isotherm $\rho_l$.

In the case that $x_{\alpha} \geq x_{\beta}$ the constant solution $\rho_l$ is not part of the density profile and the profile becomes
\begin{equation}\label{rho xw case1}
\rho(x) = \begin{cases}
\rho_{\alpha}  & \text{for $0 \leq x \leq x_w$},\\
\rho_\beta  & \text{for $ x_w \leq x \leq 1$},
\end{cases}
\end{equation}
where $x_w$ is obtained by matching the currents of the two solutions, $j_\alpha(x_w)=j_\beta(x_w)$  \citep{parmeggiani2004totally}. In this case a discontinuity appears at $x_w$.

\subsubsection*{Phase diagrams}

There are seven characteristic density profiles, called phases. Depending on $\Omega$ and $\eta$ all or some can be present in the phase diagram. We follow the same terminology as in Ref.~\citep{parmeggiani2004totally}. The simplest three are the high density (HD), the low density (LD) and the maximum current (MC) phase. In the HD (LD) phase the density is higher (lower) than $1/2$, and the density profile is given by the $\rho_\beta$ ($\rho_\alpha$) solution. In the MC phase the density profile is equal to $1/2$ over the whole domain. This is called the maximum current phase because the current is maximal for $\rho=1/2$. The density profile in the MC phase does not depend on the boundary conditions $\alpha$ and $\beta$.

Due to the interaction with the background it is possible for two or all three solutions to coexist in a density profile. This happens in the LD-HD, LD-MC, MC-HD and the LD-MC-HD phases. For example the LD-HD phase consists of the $\rho_{\alpha}$ solution on the left side and the $\rho_{\beta}$ on the right side of the lattice. The phase diagrams are constructed using the information of the domain of each of the solutions. The detailed construction of the phase diagram with $\eta=0$ is reported in Ref.~\citep{parmeggiani2004totally}.

\begin{figure}
\includegraphics[width = \columnwidth]{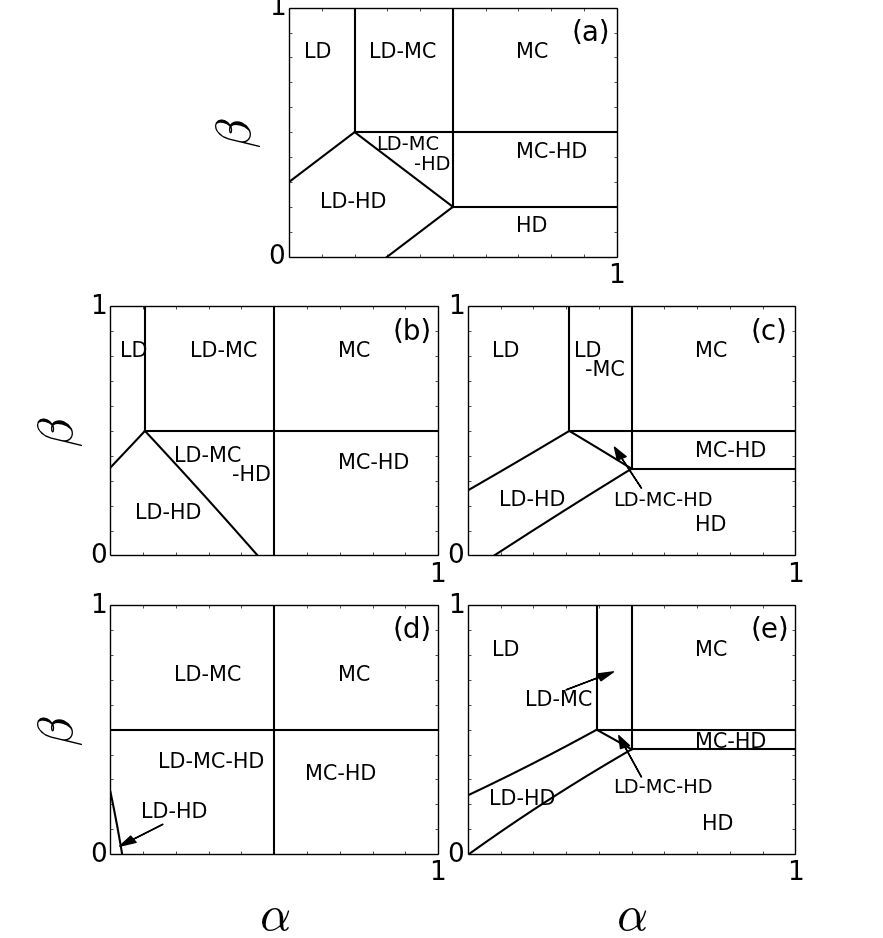}
\caption{
Five different phase diagrams for $\Omega=0.3$ and different values of $\eta$: (a) $\eta=0$, (b) $\eta=0.5$, (c) $\eta=-0.5$, (d) $\eta=2.0$ and (e) $\eta=-0.9$. (a) Corresponds to the case without mutual interaction. The influence of $\Omega$ is analysed in \citep{parmeggiani2004totally}. From (b) and (d) it becomes clear that due to the increase in $\eta$ the HD region disappears more quickly from the phase diagram than the the LD phase. This is explained by the fact that the mutual interactions are most apparent in region of high density. In cases (c) and (e)  $\eta$ is decreased. Again it becomes clear from these figures that a change in $\eta$ has more influence on the phases containing HD regions than on phases containing LD regions.
}
\label{phase_case1}
\end{figure}

The phase diagrams for nonzero $\eta$ are shown in Fig.~\ref{phase_case1} The behaviour of the phase diagrams for changes in $\eta$ are similar to changing $\Omega$. For increasing $\Omega$ the area occupied by the LD, LD-HD and the HD phase in the phase diagram decrease and and eventually disappear \citep{parmeggiani2004totally}. The key difference between changing $\Omega$ and $\eta$ is that the influence of $\eta$ is stronger in the phases containing the HD phase. As seen in Fig.~\ref{phase_case1}(b) and (d), for increasing $\eta$ the HD phase disappears quickly from the phase diagram, while the LD phases decreases slowly. This is due to the high probability of occupied neighbouring sites in regions of high density. Changes in $\Omega$ and $\eta$ do not effect the area of the MC phase, which is always confined to the upper right quarter of the phase diagram. If one keeps increasing $\eta$  eventually only the LD-MC, MC, MC-HD and LD-MC-HD phase remain in the phase diagram. Further increasing $\eta$ does not change the phase diagram any more, however the density profiles do change. For $\eta \rightarrow \infty$ the LD and HD parts of the density profile occupy an infinitely small domain on the boundaries of the profile. This means that due to the increase in Langmuir dynamics the density on the whole lattice is equal to the Langmuir density $\rho_l$.

\subsubsection*{Density profiles}
\begin{figure}
\includegraphics[width = \columnwidth]{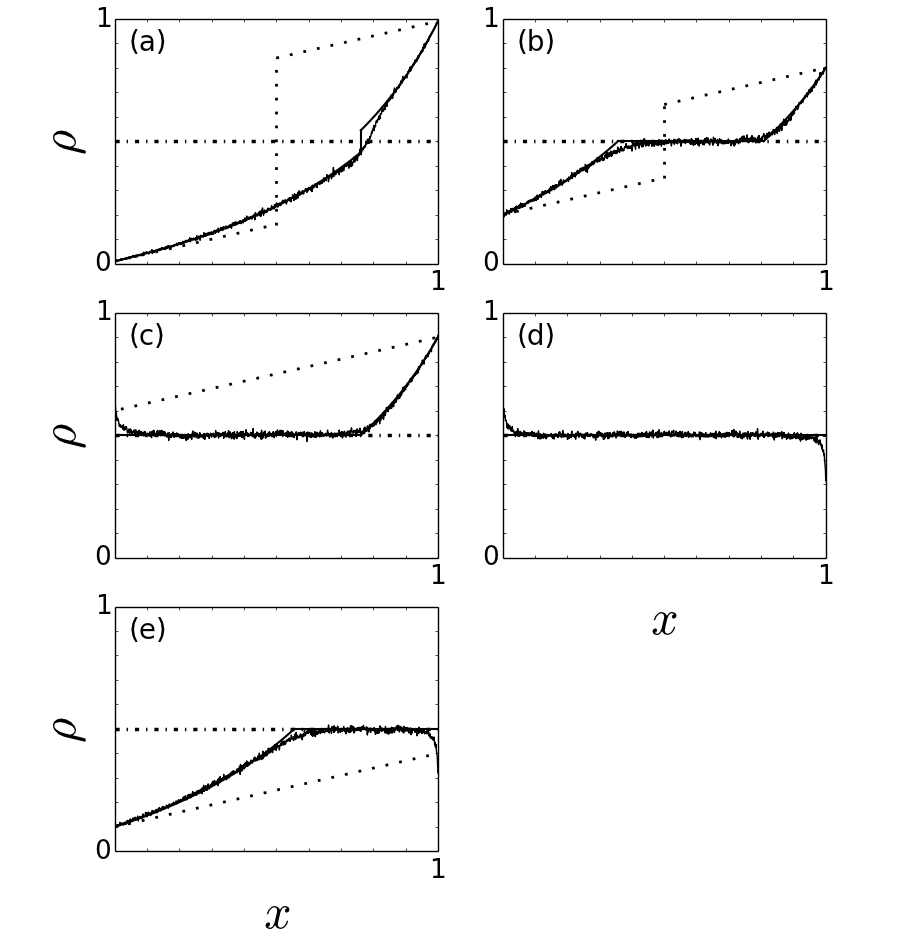}
\caption{
Density profiles for $\Omega=0.3$ and $\eta=2$, this corresponds to the phase diagram in Fig.~\ref{phase_case1} (d). The boundary conditions are (a) $\alpha=0.01$ $\beta=0.01$, (b) $\alpha=0.2$ $\beta=0.2$, (c) $\alpha=0.8$ $\beta=0.1$, (d) $\alpha=0.8$ $\beta=0.8$ and (e) $\alpha=0.1$ $\beta=0.8$. The solid lines are the analytical solutions for the density profiles (Eqs.~(\ref{rho xa xb case1},\ref{rho xw case1})), the constant solution $\rho_l$ is included as a dash-dot line. The solid lines with noise are the result of the Monte Carlo simulations with a lattice of $1000$ sites, averaged over $2000$ simulations. The analytical solutions for $\Omega=0.3$ and the same boundary conditions, but without MI ($\eta=0$) are included as a dotted line to emphasize the effect of the mutual interactions.
(a) The MI do change the density profile significantly, but do not change the phase. (b)  Due to the MI the LK dynamics increases, and the profile changes from a LD-HD phase for $\eta=0$ to a LD-MC-HD phase. (c) The MI cause a phase change from the HD phase to the MC-HD phase, due to the increased LK dynamics.(d) There is no difference between the solution with or without MI. (e) The MI induce a phase change from the LD phase to the LD-MC phase, due to the increased LK dynamics. There are two types of discrepancies between the analytical results and the Monte Carlo simulations. Boundary layers are formed at the left boundary in (c) at the right boundary in (e) and at both boundaries in (d). Other discrepancies between the analytical result and the simulations occur where there is a transition between the $\rho_\alpha$, the $\rho_\beta$ and or the $\rho_l$ solution. Both types of discrepancies can be explained by the finite size of the lattice used in the simulations.
}
\label{profile_0.3_3}
\end{figure}

With equations for $\rho_\alpha$, $\rho_\beta$ and $\rho_l$ the density profiles are constructed. In Fig.~\ref{profile_0.3_3} the density profiles for the five different phases with $\Omega=0.3$ and $\eta=2.0$ are shown ,these correspond to the phase diagram in Fig.~\ref{phase_case1}(d). The first thing to notice is that the $\rho_\alpha$ and $\rho_\beta$ solutions are not straight lines, in contrast to the solutions for $\eta=0$ which are straight lines. For $\eta>0$ the $\rho_\alpha$ and $\rho_\beta$ solutions are concave up with a positive slope. This can be explained by an increase in attraction to the Langmuir isotherm as the density increases. For example in figure \ref{profile_0.3_3} (a), if one moves away from the left boundary the density increases. This increase in density leads to an increase in mutual interactions. In the case of positive $\eta$ this results in an increase in LK dynamics and therefore an increase in attraction to the Langmuir isotherm. This increase in attraction causes the slope to increase. The  $\rho_\alpha$ and $\rho_\beta$ solutions with $\eta<0$ are concave down with positive slope, this can be explained with the same arguments as in the case of $\eta>0$.

The analytical solutions of the density profiles in the continuum limit are compared with Monte Carlo simulations of the model; see Fig.~\ref{profile_0.3_3}. Due the the finite size of the lattice used in the simulations one can expect certain discrepancies between the analytical results and the simulations.
If the $\rho_{\alpha}$ or the $\rho_{\beta}$ solutions are not part of the density profile, one or both of the boundary conditions are not met. This happens in the LD, LD-MC, MC and the MC-HD phase. In these phases a so called boundary layer forms where the boundary condition is not met \citep{parmeggiani2004totally}. A boundary layer is a discrepancy between the analytical result of the equation in the continuum limit (Eqn.~(\ref{density case1})) and the simulation results of the model with a finite sized lattice. This discrepancy occurs at the boundary where the boundary condition is not met, and will occupy an increasingly small domain for an increasing number of lattice sites used in the simulations.
One can also expect a discrepancy between the analytical density profile and the simulation results where there is a crossover from one solution to the other. In the analytical density profile the crossover is localized on the scale of the rescaled variable $x$. However if the lattice has a finite number of sites, the crossover will span a finite fraction of the normalized lattice.

\subsection*{Case 2: mutual interactions with antisymmetric modified LK rates}

In the previous case the MI increased the LK dynamics, which is not the attractive interaction as  reported in \citep{roos2008dynamic}. In this case the model is analyzed for attractive interactions. Again $\Omega_A$ and $\Omega_D$ are set equal, $\Omega_A=\Omega_D \equiv \Omega$. But in this case the mutual attraction are incorporated in an antisymmetric manner, $\delta$ is increased and $\gamma$ is decreased by a factor $\psi$, $\delta = 1+\psi$ and $\gamma = 1-\psi$, with $-1<\psi<1$. Depending on whether $\psi$ is positive or negative the interactions between the particles is respectively attractive or repulsive.
In this case Eqn.~(\ref{mi_equation_semi_continuous}) in the continuous limit becomes
\begin{equation}\label{eq_case2}
0=(2\rho-1)\partial_x \rho + \Omega[1+\rho \psi]^2 (1-\rho) \\- \Omega[1-\rho \psi]^2 \rho.
\end{equation}
In order to find solutions for the density profile, the equation is simplified by neglecting the terms of order $\psi^2$. In the next section the limit of this approximation is discussed. With this approximation Eqn.~(\ref{eq_case2}) simplifies to
\begin{equation}\label{eq_case2_approx}
0=(2\rho-1)\partial_x \rho + \Omega - 2 \Omega(1-\psi)\rho.
\end{equation}
This equation has the same form as the one for the model without MI but with unequal $\Omega_A$ and $\Omega_D$  (Eqn.~(\ref{eq:density no mi})) \citep{parmeggiani2004totally},
\begin{equation}\label{eq:density no mi}
0=(2\rho-1)\partial_x \rho +\Omega_A- (\Omega_D + \Omega_A )\rho.
\end{equation}

Eqn.~(\ref{eq:density no mi}) exhibits a particle-hole symmetry  (Eqn.~(10)), therefore this symmetry is also apparent in Eqn.~(\ref{eq_case2_approx}). However this is not a property of the model and is only apparent if the $\psi^2$ terms in Eqn.~(\ref{eq_case2}) are neglected. The transformation
\begin{subequations}\label{transformation}
\begin{align}
\rho(x) &\rightarrow 1-\rho(1-x),\\
\alpha &\rightarrow \beta,\\
\beta &\rightarrow \alpha,\\
\Omega &\rightarrow \Omega(1-2\psi),\\
\psi &\rightarrow \frac{-\psi}{1-2\psi},
\end{align}
\end{subequations}
leaves Eqn.~(\ref{eq_case2_approx}) invariant. Using this transformation density profiles for $-1<\psi<0$ can be obtained from solutions to Eqn.~(\ref{eq_case2_approx})  with $0<\psi$. Therefore the analysis is restricted to positive values of $\psi$. Solutions to Eqn.~(\ref{eq:density no mi}), obtained by \citep{parmeggiani2004totally}, are  Lambert W functions. Using these solutions one finds that the solutions to Eqn.~(\ref{eq_case2_approx}) for $\psi>0$ are
\begin{equation}\label{eq:rho_a case 2}
\rho_{\alpha}=\frac{\psi}{2(1-\psi)} \left( W_{-1} \left[-y(x) \right] + 1 \right) + \frac{1}{2} \text{\ \ for $\alpha<1/2$},
\end{equation}
\begin{equation}\label{eq:rho_b case 2}
\\ \rho_{\beta} = \begin{cases}
\frac{\psi}{2(1-\psi)} \left( W_0 \left[y(x) \right] + 1 \right) + \frac{1}{2} & \text{for $1-\beta \geq \rho_l$},\\
\frac{\psi}{2(1-\psi)} \left( W_0 \left[-y(x) \right] + 1 \right) + \frac{1}{2} & \text{for $\frac{1}{2} \leq 1-\beta \leq \rho_l$},\\
\end{cases}
\end{equation}
where $\rho_{\alpha}$ obeys the left and $\rho_{\beta}$ the right boundary condition. $y(x)$ Is given by
\begin{multline}\label{y(x)}
y(x)=\left |\frac{1-\psi}{\psi} (2\rho_0-1) -1\right| \\exp \left[2\Omega \frac{(1-\psi)^2}{\psi} (x-x_0)+\frac{1-\psi}{\psi} (2\rho_0-1) -1\right],
\end{multline}
with $\rho_0=\alpha$, $x_0=0$ for $\rho_{\alpha}$ and $\rho_0=1-\beta$, $x_0=1$ for $\rho_{\beta}$. The constant solution $\rho_l=\frac{1}{2(1-\psi)}$ is the equivalent of the Langmuir isotherm in the case without MI. The density profile is "attracted" to this constant solution, as explained in the previous section.
The solution $\rho_{\alpha}$ is stable only for $\alpha<1/2$, and $\rho_{\beta}$ for $\beta \leq 1/2$ \citep{parmeggiani2004totally}.

The full density profile is constructed from the solutions obeying the left and right boundary conditions, and calculating $x_w$, the position of the transition from $\rho_\alpha$ to $\rho_\beta$, by matching the currents of these solutions.

\subsubsection*{Phase diagrams}
\begin{figure}
\includegraphics[width = \columnwidth]{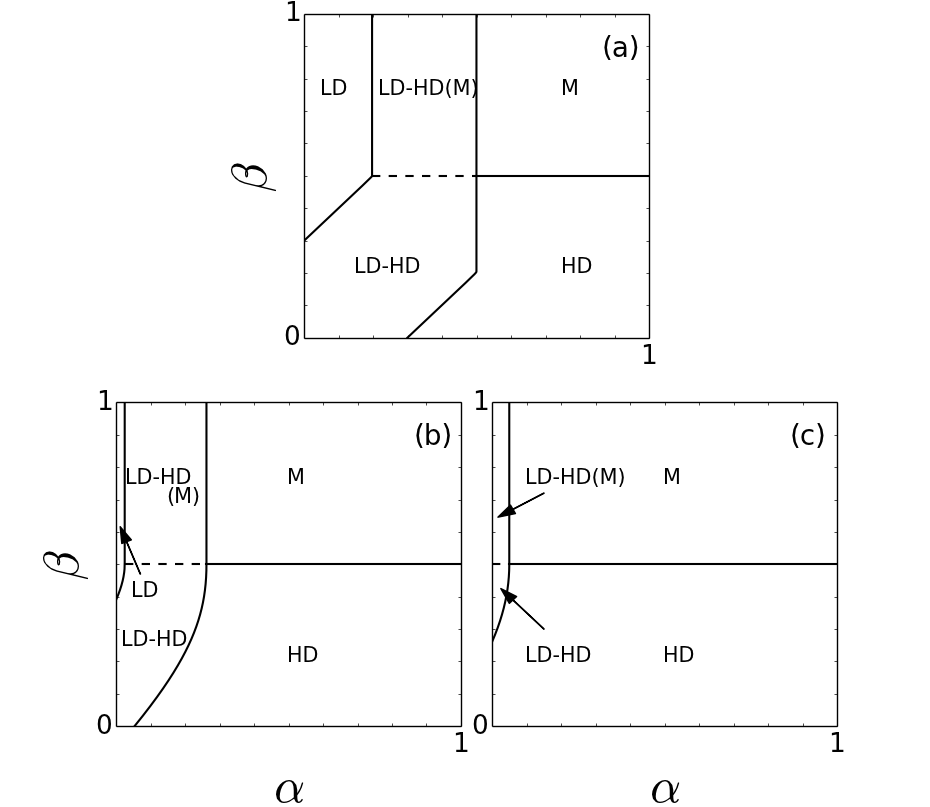}
\caption{
Three phase diagrams obtained with Eqs.~(\ref{eq:rho_a case 2}) and (\ref{eq:rho_b case 2}). With $\Omega=0.3$ and (a) $\psi=0.001 $, (b) $\psi=0.4$, (c) $\psi=0.8$. The phases which contain the $\rho_\alpha$ solution disappear from the phase diagram for increasing MI, until only the M and HD phases are left. The area of the M phase in the phase diagram occupies an increasingly large part of the upper half of the phase diagram for increasing $\psi$. This stands in contrast with the MC phase in case 1, where the MC phase is confined to the upper right quarter of the phase diagram and the area is independent of any parameter; see Fig.~\ref{phase_case1}.
}
\label{case2_phase_diagram}
\end{figure}

Using the same method as in the previous case the phase diagrams are constructed from the information about the domain of the solutions. There are four possible phases, these have the same characteristics as the phases of the model without MI but with $\Omega_A \neq \Omega_D$ \citep{parmeggiani2004totally}. Due to the similarity only a short explanation is given here, for a more elaborate discussion of the phases one can consult \citep{parmeggiani2004totally}.
In the LD (HD) phase the full density profile is governed by $\rho_\alpha$ ($\rho_\beta$); boundary layers appear  at the right (left) end of the lattice. The condition for the LD phase is $x_w>1$ and $\alpha<1/2$, and for the HD phase the condition is $x_w<0$ and $\beta<1/2$.  The M phase occurs for $\beta>1/2$ and $x_w<0$.
 This phase is called the "Meissner" phase due to similarities with the Meissner phase in super conducting materials \citep{parmeggiani2004totally}. In the M phase the density profile is independent of both boundary conditions, therefore boundary layers occur at both ends of the lattice. Because the solution is not stable for these values of $\beta$ the profile is given by  $\rho_\beta(1/2)$ \citep{parmeggiani2004totally}. This phase can be seen as the equivalent of the MC phase in case without MI \citep{parmeggiani2004totally} or case 1, because a profile in the MC phase is also independent of the boundary conditions.
The LD-HD phase, where phase coexistence occurs,  is split in two regions. In the region $\beta<1/2$ the profile obeys both the left and right boundary condition. In the region $\beta>1/2$ only the left boundary condition is obeyed. The right part of the density profile is independent of the right boundary condition and is given by $\rho_\beta(1/2)$, this phase is indicated as LD-HD(M). Profiles in the LD-HD(M) phase have a boundary layer at the right end of the lattice. The conditions for the LD-HD phase are $0<x_w<1$, $\alpha<1/2$ and $\beta<1/2$. For the LD-HD(M) phase the conditions are $0<x_w<1$, $\alpha<1/2$ and $\beta>1/2$. In Fig.~\ref{case2_phase_diagram} three phase diagrams are shown for different values of $\psi$.

Because Eqs.~(\ref{eq:rho_a case 2}) and (\ref{eq:rho_b case 2}) are derived using the approximation $\psi^2=0$, the phase diagrams are also an approximation which hold in the limit of small $\psi$.

\subsubsection*{Density profiles}
\begin{figure}
\includegraphics[width = \columnwidth]{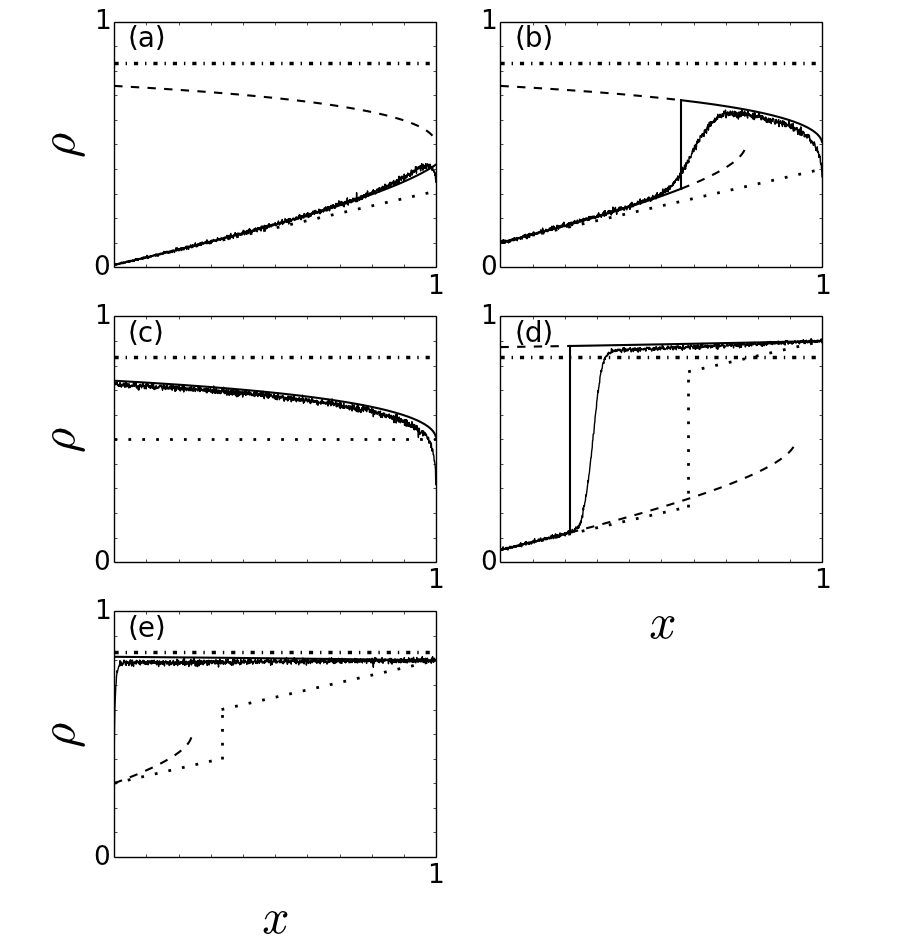}
\caption{Density profiles for $\Omega=0.3$ and $\psi=0.4$, this corresponds to the phase diagram in Fig.~\ref{case2_phase_diagram} b. The injection/extraction rates are (a) $\alpha=$0.01, $\beta=0.7$ (LD phase), (b) $\alpha=0.1$, $\beta=0.7$ (LD-HD(M) phase), (c) $\alpha=0.8$, $\beta=0.8$ (M phase), (d) $\alpha=0.05$, $\beta=0.1$ (LD-HD phase) and (e) $\alpha=0.3$, $\beta=0.2$ (HD phase). The dashed lines are the Lambert $W$ function solutions to Eqn.~(\ref{eq_case2_approx}) obeying the left and right boundary conditions. The parts of these solutions that make up the density profile are represented as solid black lines. The dash-dot line is the constant solution. The analytical solutions for the same values but without MI are included as dotted lines  to emphasize the influence of the MI on the density profiles. Solid lines with noise are the results of the Monte simulations with a lattice of $1000$ sites, averaged over $2000$ simulations.
Over all there is good agreement between the simulations and the analytical result. There are two causes for the discrepancies, the finite size of the lattice and the approximation $\psi^2 \approx 0$. Due to the finite size of the lattice boundary layers are formed at the right end of (a),(b),(d) and at the left end of (c) and (d); and the domain walls in (b) and (d) are not localized.  The discrepancies caused by the approximation are visible in two ways, the density profile does not fully coincide with the analytical result and due to this the position of the domain wall is shifted. This is visible in (b) and (d).
}
\label{case2_profiles}
\end{figure}

Using Eqs.~(\ref{eq:rho_a case 2}) and (\ref{eq:rho_b case 2}) the density profiles can be constructed. The domain of each of the solutions is determined by matching the currents of the solutions, Eqn.~(\ref{curent}). In Fig.~\ref{case2_profiles},  five density profiles are depicted, one each for a phase in phase diagram \ref{case2_phase_diagram} (b) ($\Omega=0.3,\psi=0.4$). It is clear from the Fig.~\ref{case2_profiles} that there is good agreement between the simulations and the analytical solutions (Eqs.~(\ref{eq:rho_a case 2},\ref{eq:rho_b case 2})). The apparent discrepancies between the analytical and numerical results are caused by the finite size of the lattice used in the simulations. Besides this there is also a small discrepancy between the analytical solution and the simulations caused by the approximation $\psi^2=0$. This can cause a discrepancy in the domain wall position, as can be seen in Fig.~\ref{case2_profiles}(b) and (d).

\subsubsection*{Approximation limits}
\begin{figure}
\includegraphics[width = \columnwidth]{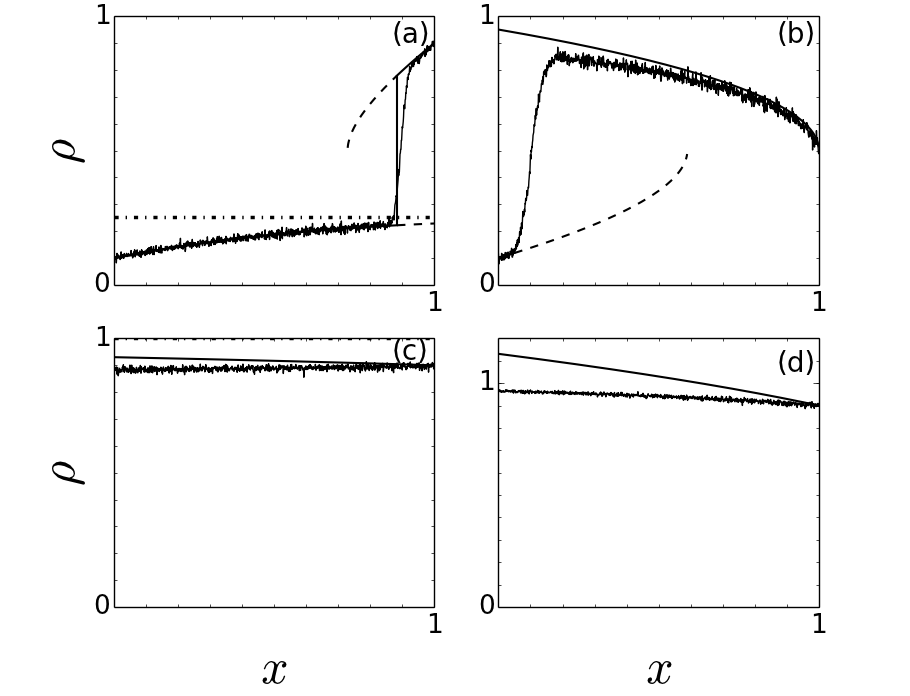}
\caption{The same legend as in Fig.~\ref{case2_profiles} is used. In order to illustrate the limits of the approximation $\psi^2 \approx 0$ used in deriving the equations for the density profiles, four extreme cases are shown. For all figures $\Omega=0.3$ was used an for (a) $\psi=-0.99,\ \alpha=0.1,\ \beta=0.1$, (b) $\psi=0.8,\  \alpha=0.1,\ \beta=0.9$, (c) $\psi=0.5,\ \alpha=0.9,\ \beta=0.1$ and for (d) $\psi=0.9,\ \alpha=0.9,\ \beta=0.1$. From (a) it is clear that the approximation holds for low densities, regardless of the values of $\psi$. The analytical solution in (b) does not fully coincide with the simulations. There are two causes for the discrepancies. First of all there is a boundary layer on the right side caused by the finite size of the lattice. Secondly  the $\rho_\beta$ solution is higher than the results from the simulation, this is caused by neglecting the $\psi^2$ terms. Over all the simulations are in good agreement with the analytical solutions, however the domain of the solutions differs significantly. The analytical profile is in the HD phase, $x_w<0$. The simulation result, on the other hand, is in the LD-HD phase, $0<x_w<1$. (c) For this value of $\psi$ there is agreement between the analytical solution and the simulations, even though density is high. (d) The analytical result does not coincide with the simulations due to the combination of high density and a $\psi$ close to 1. In this case the density of the analytical result exceeds 1, which is physically impossible.
}
\label{case2_approx}
\end{figure}

In deriving Eqs.~(\ref{eq:rho_a case 2}) and (\ref{eq:rho_b case 2}) terms of the order $\psi^2$ were neglected. The neglected part of Eqn.~(\ref{eq_case2}) is $\Omega \psi^2 \rho^2 - 2 \Omega \psi^2 \rho^3$, which shows that every $\psi^2$ is coupled to either a $\rho^2$ or a $\rho^3$. This means that the break down of the approximation is governed by both $\rho$ and $\psi$ and the approximation hold for low densities regardless of the value of $\psi$, because MI do not play a significant role in low densities due to the low probability of having occupied neighbouring sites. Fig.~\ref{case2_approx} illustrates some of the limits of the approximation. From Figs.~\ref{case2_profiles}(b),(d) and \ref{case2_approx} (b) it becomes clear that the domain of the low and high density solution can differ significantly even if the $\rho_\alpha$ and $\rho_\beta$ differ only little from the simulation result.

\subsection*{Case 3: mutual interactions with arbitrarily modified LK rates}
\begin{figure}
\includegraphics[width = \columnwidth]{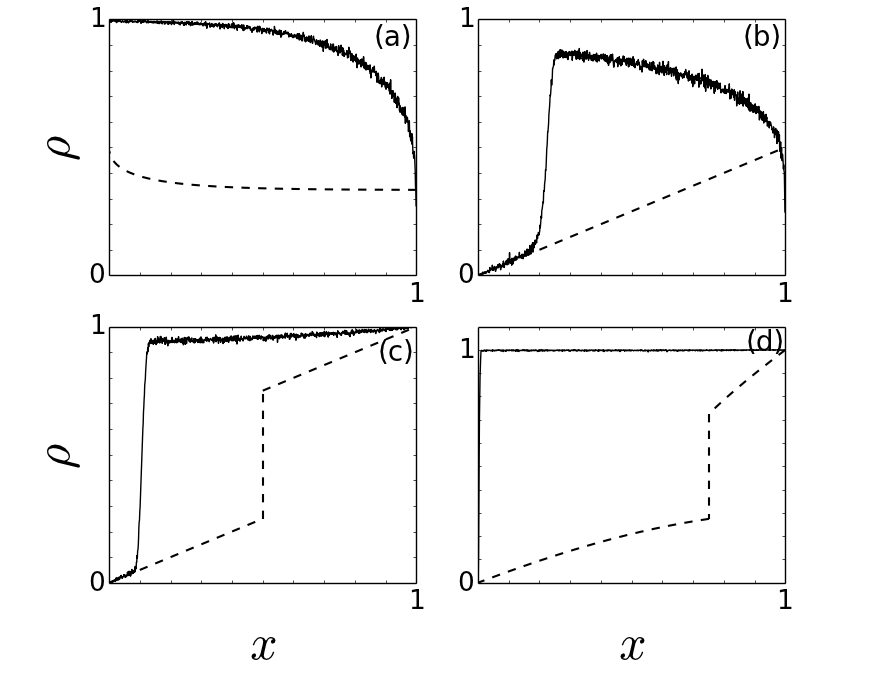}
\caption{The density profiles for (a) $\Omega_A= 0.5$, $\Omega_D=1$, $\delta=2$, $\gamma=0.1$, $\alpha=1$ and $\beta=1$, (b) $\Omega_A= 0.5$, $\Omega_D=0.5$, $\delta=2$, $\gamma=0.5$, $\alpha=0$ and $\beta=1$, (c) $\Omega_A= 0.5$, $\Omega_D=0.5$, $\delta=2$, $\gamma=0.5$, $\alpha=0$ and $\beta=0$, (d) $\Omega_A= 0.5$, $\Omega_D=1$, $\delta=3$, $\gamma=0.1$, $\alpha=0$ and $\beta=0$. Solid line with noise is the result from the simulation, the dashed line is the solution without MI. In all cases the attachment/detachment is increased/decreased, which results in a higher density. In (b) and (c) the density overlaps with a small part of the analytical solution without MI. This can be explained by the in low density regions the effect of the MI is small. Though in (d) the solution without MI the density is low, the lattice is almost completely filled due to the increase/decrease in attachment/detachment. All parameters in (a) and (d) are the same except for the boundary conditions, which prevents the lattice in (a) from filling up completely.}
\label{case3_profiles}
\end{figure}

Until now, we have considered enhancement or suppression of LK rates in a symmetric or antisymmetric fashion. The most general case in which all the relevant parameters ($\Omega_A$, $\Omega_D$, $\delta$, $\gamma$, $\alpha$, $\beta$) are assigned randomly chosen values is extremely difficult to treat analytically. Due to the large parameter-space (6-dimensional), one cannot gain much insight by performing numerical simulations. Therefore, we have not explored the most general case in any details. However, we consider few representative cases in which the choice of model parameters is based on the observation that the resulting density profiles have interesting features when contrasted with the case with no mutual interactions. In Fig.~\ref{case3_profiles} we show profiles corresponding to four different sets of parameters. As can be seen the profiles look very different from those with no mutual interaction. However, as mentioned above, at present our analysis of the most general case is very qualitative and highly superficial. It is obvious that it requires much further investigation. We leave detailed analysis of the general case to a future study.


\section{conclusion}\label{conclusion}
In this work, we investigate the simple one-dimensional driven model-- the totally asymmetric exclusion process, coupled to a modified form of Langmuir kinetics. This model is motivated by recent studies on clustering of motor proteins on microtubules. Without addressing the underlying mechanism, it is assumed that the attachment and detachment rates of a particle depend on the occupancy of the nearest neighbours of any given site. Ignoring density correlations, we obtain continuum mean-field equation describing the density profile on the lattice. Imposing certain conditions, we obtain analytical solution to the equation and demonstrate using Monte Carlo simulations that our analytical solutions are accurate over a wide range of parameters. We show that when both attachment and detachment rates are enhanced due to mutual interactions, bulk-effects start dominating the phase behavior of the model. The two-phase coexistence (low and high density) observed in absence of mutual interactions can become three-phase coexistence (low and high density with maximum current phase) when mutual interactions are attractive. On varying the mutual interaction between particles (attractive or repulsive), we obtain a very rich phase diagram describing the behavior of driven diffusive system with mutually interactive Langmuir kinetics.

%

\bibliographystyle{apsrev}
\bibliography{Bibliography}

\end{document}